\documentclass[aps,pra,amssymb,twocolumn, amsfonts,amsmath,showpacs, superscriptaddress,nofootinbib]{revtex4}
 \usepackage{graphicx}
  \usepackage{amssymb}
 
\begin{document}
\title{Collapse Miscellany}
\author{Philip Pearle}
\affiliation{Emeritus, Department of Physics, Hamilton College, Clinton, NY  13323}
\begin{abstract}
 {An introduction to the CSL (Continuous Spontaneous Localization) theory of dynamical wave function 
 collapse is provided, including a derivation of CSL from two postulates. There follows   applications  
to a free particle, or to a `small'  rigid cluster of free particles, in a single wave-packet and in interfering packets.}
 \end{abstract}
 \email{ppearle@hamilton.edu}
\maketitle

\section{Introduction}\label{sec1} 
Standard quantum  theory is readily applied to measurement situations. This requires additional (ad hoc--for this case only) information be supplied  
for each situation.  One may hope the theory could be extended to describe reality independent of experiment and without the need for ad hoc information. 
Given a state vector for any physical system, no matter how large or complex, one may hope for a theory that specifies   
the state vectors corresponding to the possible realizable states of nature, and their probabilities of realization. 

Why can't this be done with standard quantum theory?  Given a state vector, 
\begin{equation}\label{1}
|\psi,t\rangle=\sum_{n}|a_{n}\rangle\langle a_{n}|\psi,t\rangle,
\end{equation}
\noindent one would like to say that the $|a_{n}\rangle$ correspond to the possible realizable states of nature, and that $|\langle a_{n}|\psi,t\rangle|^{2}$ are their probabilities of 
realization. 

There are two problems with this. The first is the \textit{preferred basis problem}. Why $|a_{n}\rangle$ and not another orthonormal basis $|b_{n}\rangle$? 
No one has been able to specify the needed preferred basis. Of course, in experimental situations and special model situations this is possible, but it has not been possible in general. 

The second I like to call the \textit{hopping problem}.  If $|\langle a_{n}|\psi,t\rangle|^{2}$ is the probability of $|a_{n}\rangle$ being realized in nature at time $t$, and 
$|\langle a_{m}|\psi,t\rangle|^{2}$ is the probability of $|a_{m}\rangle$ ($m\neq n$) being realized in nature at time $t+dt$, then 
\[
|\langle a_{n}|\psi,t\rangle|^{2}|\langle a_{m}|\psi,t+dt\rangle|^{2}
\]
\noindent is the probability of occurrence of both these events which, of course, is not seen.  The difficulty is that quantum theory just gives the probabilities of these events:  
it does not give the probability of transitions between these events.

The resolution chosen by the Founding Fathers  was to restrict quantum theory to experimental situations, with  
empirically defined preferred bases, and to adopt the `collapse postulate,' to ensure that a preferred basis state, once chosen by nature, would remain chosen.
They thereby relinquished the hope to describe reality.  

An alternative resolution, described here, does not give up on this hope. It alters quantum theory so that it may be realized.  The Schr\"odinger 
equation is modified by adding a randomly fluctuating term to the Hamiltonian, to account for the probabilistic behavior of nature. There is a preferred basis built into this term.   The state vector dynamically collapses toward one of these basis states, very slowly for micro-objects, very rapidly 
for macro-objects.  

The preferred basis is essentially the mass density basis.  This choice could not be made for standard quantum theory, since  collapse to position eigenstates gives the particles infinite energy.  It works here because the collapse never goes all the way.  First, it  would take 
infinite time for that to occur.  Second,  it  doesn't happen because of the interaction of the usual Hamiltonian dynamics with the collapse dynamics (e.g.,  the example in Section \ref{SP}). 
\section{Deriving CSL}\label{sec2}

For collapse dynamics, one wants the following behavior. 

Given the state vector Eq.(\ref{1}) at time  $t=0$, the squared amplitudes $x_{n}(t)\equiv|\langle a_{n}|\psi,t\rangle|^{2}$ should fluctuate until eventually one amplitude becomes equal to 1, and the rest equal to 0: that is the collapse. 

 Moreover, with repeated evolutions,  the $n$th amplitude should eventually reach 1 for a fraction $x_{n}(0)$ of the evolutions: that is the Born rule.

\subsection{Gambler's Ruin Game}\label{sec2A} 
There is a rather precise analogy to this behavior,  the `gambler's ruin game.'\cite{gambler}  Consider two gamblers,  with \$100 between them. Gambler 1 starts with \$$X_{1}(0)$, gambler 2 with  \$$X_{2}(0)=$ \$$100-X_{1}(0)$.  They toss a fair coin: `fair' is crucial, making the game what mathematicians call a 'Martingale.'  Heads, gambler 1 gives a dollar to gambler 2, tails, the reverse.  The amount each possesses, $X_{n}(t_{k})$, fluctuates.  Eventually, the game ends, with one gambler in possession of all the money.   

Define $P(X_{1})$ as the conditional probability that gambler 1 eventually wins, given that he has \$$X_{1}$.  Then, 
\[
P(X_{1})=\frac{1}{2}P(X_{1}-1)+\frac{1}{2}P(X_{1}+1).
\]
\noindent That is, when the coin is tossed, gambler 1 can either lose the toss but win from there, or win the toss and win from there.  The solution of this difference equation is
 $P(X_{1})=AX_{1}+B$ ($A$ and $B$ are constants).  With boundary conditions $P(0)=0$ and $P(100)=1$, the solution is $P(X_{1})=X_{1}/100$. 
  
 Define $x_{n}(t_{k})\equiv X_{n}(t_{k})/100$.  Thus, we have seen that  if gambler 1 starts out with a fraction $x_{1}(0) $ of the total amount of money,  that is  the fraction of repeated games he wins.    
 
 This is just the behavior we want for a state vector which is the sum of  two basis vectors $|a_{1}\rangle$, $|a_{2}\rangle$.  Thus, the two basis vectors may be thought of as competing in a continuous version of the gambler's ruin game.  
  
For,  the gambler's ruin game is a zero-sum,  discrete, fair, random walk with absorbing barriers (at $x_{i}=0,1$)  in discrete time, for 
$x_{n}(t_{k})\equiv X_{n}(t_{k})/100$. 

 Collapse  is a zero-sum, continuous, fair, random walk with absorbing barriers in continuous time, for $x_{n}(t)\equiv|\langle a_{n}|\psi,t\rangle|^{2}$.  
   
The continuous limit of discrete random walk is Brownian motion.  Thus, it is natural to regard the  $x_{n}(t)$ as undergoing some kind of Brownian motion.  We shall replace the  probabilities  associated with a single coin toss, and  with a sequence of  coin tosses respectively by 
\begin{subequations}
\begin{eqnarray}\label{2}
P(dB)&\equiv&\frac{1}{\sqrt{2\pi \lambda dt}}e^{-dB^{2}/2\lambda dt},\label{2a}\\
P(B(t))&\equiv&\frac{1}{\sqrt{2\pi \lambda t}}e^{-B(t)^{2}/2\lambda t}\label{2b}
\end{eqnarray}
\end{subequations}
 \noindent where $\lambda$ is a constant diffusion rate. Just as the gambler's ruin dollar count $X_{n}(t_{k})$ depends upon a sequence of coin tosses, so 
  we shall take the collapse $x_{n}(t)$ to depend upon $B(t)$. 
  
  \subsection{Postulates}\label{sec2B} 
  
  We shall derive the CSL collapse dynamics from two postulates:  
 \begin{eqnarray}
&&\hbox{ 1) gambler's ruin behavior.}\nonumber\\
&&\hbox{ 2) a linear, real, Schr\"odinger 
  equation.}\nonumber
\end{eqnarray}
   
  2) could use some explanatory remarks.
 
  In order to achieve the \textit {linearity} part of 2), it shall be necessary to relinquish the condition that the state vector norm is 1.  In  standard 
  quantum theory, the unit norm condition is mandatory, since it ensures that probabilities add up to 1.  Here there is no need for the unit norm condition: probabilities are 
  provided by $B(t)$'s probabilities.   Physical information is carried by a state vector's direction in Hilbert space, not its norm. To calculate expectation values, one can always normalize a state vector.  The un-normalized state vector shall be denoted $|\phi,t\rangle$.
  
Regarding the \textit{real} part of 2), we wish to consider a different dynamics from the usual Hamiiltonian dynamics $d|\psi,t\rangle=-iHdt|\psi,t\rangle$.   That utilizes a hermitian hamiltonian $H$, so we shall consider dynamics of the form  $d|\psi,t\rangle=H_{c}dt|\psi,t\rangle$ where the `collapse hamiltonian' $H_{c}$ is hermitian  (we note that the most general linear equation is of the form $d|\psi,t\rangle=-i(H+iH_{c})dt|\psi,t\rangle$).  

Now, for the purposes of this derivation, for simplicity and appropriate to the generalization of a classical game, we shall further restrict the state vector components to be real numbers. Therefore, we must restrict $H_{c}$  to be a real symmetric operator in the chosen basis $|a_{n}\rangle$. Once we find the unique form of  $H_{c}$  that allows postulate 1), it may readily be seen that that complex state vector components and  the most general hermitian $H_{c}$  also allow postulate 1).  Because the state vector components are real, we  may write 
  \begin{equation}\label{3}
  x_{n}(t)=\frac{\langle a_{n}|\phi,t\rangle^{2}}{\langle\phi,t|\phi,t\rangle}=\frac{\langle a_{n}|\phi,t\rangle^{2}}{\sum_{m}\langle a_{m}|\phi,t\rangle^{2}}.
\end{equation}
  
  Also in order to achieve the linearity part of 2), the Schr\"odinger equation shall be linear in a Brownian motion $B'(t)$. However, the relation assumed between 
  $B'(t)$ and $B(t)$ is  allowed to be non-linear in the state vector: 
  
  \begin{equation}\label{4}
dB'(t)=dB(t)+f({\bf x})dt. 
\end{equation}

\noindent where $f$ is an arbitrary real function of the $x_{n}(t)$.  For, collapse violates  the superposition principle, and this \textit{somehow} requires a non-linearity.  What 2) really means, then, is that we look to have a linear Schr\"odinger equation and isolate all the non-linearity in the probability.\cite{Pearle89}
    
  The derivation, which is presented in the next few sections, is rather lengthy, and it involves stochastic differential equations--but the result does not.  Some readers may wish to move immediately to  section \ref{secIII} where the results obtained here are utilized. 
  
  Postulate 1) is implemented by an \textit{It\^o} equation for $x_{n}(t)$:
  \begin{equation}\label{5}
dx_{n}(t)=b_{n}({\bf x})dB(t),
\end{equation}
\noindent where $b_{n}$ is an arbitrary real function of the $x_{m}(t)$'s.  

We shall denote by an overline the ensemble average of a quantity, e.g., from Eq.(\ref{2a}), $\overline{dB}=\int d(dB)P(dB)=0$. 

  It immediately  follows from Eq.(\ref{5}) that  $\overline{dx_{n}}(t)=0$. This says that it is a `fair game' for each `player'
which, as we have said, is crucial for gambler's ruin behavior.\cite{Pearle76}  This is why Eq.(\ref{5}) is chosen to be an It\^o equation.

The other necessary condition  for gambler's ruin behavior is the end-game condition, that $b_{n}({\bf x})=0$ when one $x_{n}(t)$ is equal to 1 and the rest vanish.  However, that does not need to be separately imposed since it automatically occurs, as shall be seen.

Postulate 2) is  implemented by the \textit{Stratonovich} Schr\"odinger equation for the un-normalized state vector amplitudes: 
\[
d|\phi,t\rangle=[RdB'+Sdt]|\phi,t\rangle
\]
\noindent where $R$ and $S$ are arbitrary symmetric real operators, as discussed above. This is chosen to be a Stratonovich equation because the calculus manipulations (e.g., derivative of the product of functions)  are the usual ones, which would not be the case for an It\^o equation.  

 However, a rather tedious calculation (Appendix A) shows that  
$R$ and $S$ have to be diagonal in the $|a_{n}\rangle$ basis if postulate 2) is to imply postulate 1).  Therefore, we shall write this equation as

 \begin{equation}\label{6}
d\langle a_{n}|\phi,t\rangle=[\alpha_{n}dB'+\beta_{n}dt]\langle a_{n}|\phi,t\rangle
\end{equation}
\noindent where $\alpha_{n}, \beta_{n}$ are real constants.   

\subsection{Derivation}\label{sec2C} 

We proceed to find $d{x_{n}}(t)$ from Eq.(\ref{3}).  With use of Eqs.(\ref{4}),(\ref{6}), we obtain  the Stratonovich equation 
\begin{eqnarray}\label{7}
d{x_{n}}(t)&=&2\{(\alpha_{n}-{\bf \alpha}\cdot{\bf x})dB\nonumber\\
&+&[(\alpha_{n}-{\bf \alpha}\cdot{\bf x})f+(\beta_{n}-{\bf\beta}\cdot{\bf x})]dt\}x_{n}(t). 
\end{eqnarray}
\noindent where  ${\bf\alpha}\cdot{\bf x}\equiv \sum_{m}\alpha_{m}x_{m}(t)$

Now we may use the rule for converting a Stratonovich equation to an It\^o equation, which in this case means adding 
\[
\frac{\lambda dt}{2}\sum_{m}2\{(\alpha_{m}-{\bf \alpha}\cdot{\bf x})\frac{\partial}{\partial x_{m}}2\{(\alpha_{n}-{\bf \alpha}\cdot{\bf x})
\]
\noindent to the right side of Eq.(\ref{7}). The result is
\begin{eqnarray}\label{8}
&&d{x_{n}}(t)=2\{(\alpha_{n}-{\bf \alpha}\cdot{\bf x})dB\nonumber\\
&+&[(\alpha_{n}-{\bf \alpha}\cdot{\bf x})f+(\beta_{n}-{\bf\beta}\cdot{\bf x})]dt\nonumber\\
&+&[-{\bf\alpha}^{2}\cdot{\bf x}+({\bf \alpha}\cdot{\bf x})^{2}+(\alpha_{n}-{\bf \alpha}\cdot{\bf x})^{2}]\lambda dt\}x_{n}(t).
\end{eqnarray}
\noindent  where $ {\bf\alpha}^{2}\cdot{\bf x}\equiv\sum_{m} \alpha_{m}^{2}x_{m}(t)$.  In order that the It\^o Eq.(\ref{8}) (consequence of postulate 2) agree with the It\^o Eq.(\ref{5}) (consequence of postulate 1), the coefficient of $dB$ and the coefficient of $dt$ in both equations must be equal, so
\begin{subequations}
\begin{eqnarray}\label{9}
&&b_{n}({\bf x})=2(\alpha_{n}-{\bf \alpha}\cdot{\bf x})x_{n}(t),\label{9a}\\
&&(\alpha_{n}-{\bf \alpha}\cdot{\bf x})f+(\beta_{n}-{\bf\beta}\cdot{\bf x})=\nonumber\\ 
&&\qquad\quad\lambda [{\bf\alpha}^{2}\cdot{\bf x}-({\bf \alpha}\cdot{\bf x})^{2}-(\alpha_{n}-{\bf \alpha}\cdot{\bf x})^{2}]. \label{9b}                                                                                                   
\end{eqnarray}
\end{subequations}

To find $f$ and $\beta_{n}$, operate on  Eq.(\ref{9b}) with $\sum_{n}x_{n}\partial^{2}/\partial x_{m}^{2}$.  Remembering that $\sum_{n}x_{n}=1$, we obtain
\begin{equation}\label{10}
2\alpha_{m}\frac{\partial}{\partial x_{m}}f=4\lambda\alpha_{m}^{2}, \hbox{  or  } f=2\lambda {\bf \alpha}\cdot{\bf x}+c.
\end{equation}
\noindent where $c$ is an arbitrary constant.  Putting Eq.(\ref{10}) back into Eq.(\ref{9b}) gives
\begin{equation}\label{11}
(\beta_{n}-{\bf\beta}\cdot{\bf x})=-c(\alpha_{n}-{\bf \alpha}\cdot{\bf x})-\lambda(\alpha_{n}^{2}-{\bf\alpha}^{2}\cdot{\bf x}), \hbox{  or  }\beta_{n}=-c
\alpha_{n}-\lambda\alpha_{n}^{2}, 
\end{equation}
\noindent the last step following from applying $\partial/\partial x_{i}$ to the first.  Putting $f$ from Eq.(\ref{10}) and $\beta_{n}$ from Eq.(\ref{11}) into  
Eq.(\ref{7}), we find that the latter is independent of $c$. 
Therefore, $c$ has no physical effect.  Its only effect is to add a constant drift to $B'(t)$ (see  Eq.(\ref{5})), and since we are free to choose $B'(t)$ to be as 
simple as possible, we may take $c=0$.  

	This concludes our derivation.  We have found that postulates 1) and 2) lead to the \textit{Schr\"odinger equation} (\ref{6}) and the \textit{Probability 
Rule} (\ref{4}), in the forms

\begin{subequations}
\begin{eqnarray}\label{12}
d\langle a_{n}|\phi,t\rangle&=&[\alpha_{n}dB'-\lambda\alpha_{n}^{2}dt]\langle a_{n}|\phi,t\rangle,\label{12a}\\
dB'(t)&=&dB(t)+2\lambda {\bf \alpha}\cdot{\bf x}dt, \label{12b}                                                                                                   
\end{eqnarray}
\end{subequations}
\noindent where the $\alpha_{n}$ are completely arbitrary constants and $\bf{x}$ is given by Eq.(\ref{3}). 
\subsection{Schr\"odinger Equation}\label{sec2D}

We may introduce an operator $A$ defined by $A|a_{n}\rangle=\alpha_{n}|a_{n}\rangle$, in 
terms of which the Schr\"odinger equation (\ref{12}) may be written in basis-independent form, 
\begin{equation}\label{13}
d|\phi,t\rangle=[AdB'-\lambda A^{2}dt]|\phi,t\rangle \hbox  { or } |\phi,t\rangle=e^{AB'(t)-A^{2}\lambda t}|\phi,0\rangle.
\end{equation}
\noindent  Since the $|a_{n}\rangle$ are eigenstates of $A$ with eigenvalues $\alpha_{n}$, we could instead label them $|\alpha_{n}\rangle$.  However, we shall rather replace $\alpha_{n}$ by  $a_{n}$. 
\subsection{Probability Rule}\label{sec2E} 

Eq.(\ref{12b}) is the same form whether an It\^o or Stratonovich equation. It may be written as
\begin{equation}\label{14}
dB'(t)=dB(t)+2\lambda dt\langle A\rangle(t)\hbox{ where }\langle A\rangle(t) \equiv\frac{\langle \phi,t|A|\phi,t\rangle}{\langle \phi,t|\phi,t\rangle}.
\end{equation}
\noindent We shall use it to find the expression for the probability of $B'(t)$.  

It follows from Eq.(\ref{14}) that $\overline{dB'(t)}=2\lambda dt\langle A\rangle(t)$, 
$\overline{(dB'(t))^{2}}=\lambda dt$, and all higher moments of $dB'(t)$ can be neglected as they are of higher order than $dt$. Therefore, $dB'(t)$ is a gaussian
 process, characterized by just these two moments.  One may immediately check that the probability  density, variously written as  
 \begin{subequations}
 \begin{eqnarray}\label{15}
P[dB'(t)]&=&\frac{1}{\sqrt{2\pi\lambda dt}}e^{-\frac{(dB'(t))^{2}}{2\lambda dt}}
\frac{\langle\phi,t| e^{2AdB'(t)-2A^{2}\lambda dt} |\phi,t\rangle}{\langle\phi,t|\phi,t\rangle}\nonumber\\
&=&\frac{1}{\sqrt{2\pi\lambda dt}}\frac{\langle\phi,t|e^{-\frac{(dB'(t)-2\lambda dtA)^{2}}{2\lambda dt}} |\phi,t\rangle}{\langle\phi,t| \phi,t\rangle}\label{15a}\\
&=&\frac{1}{\sqrt{2\pi\lambda dt}}e^{-\frac{(dB'(t))^{2}}{2\lambda dt}}\frac{\langle\phi,t+dt| \phi,t+dt\rangle}{\langle\phi,t| \phi,t\rangle}\label{15b}
\end{eqnarray}
\end{subequations}
 has these two moments.
 
 Define $N\equiv t/dt$. We multiply Eq.(\ref{15b}) by itself $N+1$ times with successively smaller values of $t$, obtaining  the joint probability of the 
 independent increments $dB$ at successive values of $t$:
 \begin{eqnarray}\label{16}
\prod_{n=0}^{N}P[dB'(t-ndt)]&=&\prod_{n=0}^{N}\frac{1}{\sqrt{2\pi\lambda dt}}e^{-\frac{(dB'(t-ndt))^{2}}{2\lambda dt}}\nonumber\\
&&\qquad\cdot\frac{\langle\phi,t+dt| \phi,t+dt\rangle}{\langle\phi,0| \phi,0\rangle}.
\end{eqnarray}
\noindent Since $\langle\phi,0| \phi,0\rangle=1$, and using Eq.(\ref{13}) to write
\begin{eqnarray}\label{17}
&&\langle\phi,t+dt |\phi,t+dt\rangle=\langle\phi,0|e^{2AB'(t+dt)-2A^{2}\lambda (t+dt)}|\phi,0\rangle\nonumber\\
&&\negmedspace\negmedspace\negmedspace
=\langle\phi,0|\prod_{n=0}^{N}e^{2A(B'(t+dt-ndt)-B'(t-ndt)-2A^{2}\lambda dt)}|\phi,0\rangle
\end{eqnarray}
\noindent (taking $B(0)=0$), it follows from Eqs.(\ref{16}),(\ref{17}) that the joint probability of the values of $B$ at successive values of $t$ is 
\begin{eqnarray}\label{18}
&&P[B'(t+dt),B'(t),..,B'(dt) ]=\prod_{n=0}^{N}\frac{1}{\sqrt{2\pi\lambda dt}}\nonumber\\
&&\qquad\cdot\langle\phi,0|e^{-\frac{(B'(t+dt-ndt)-B'(t-ndt)-2\lambda dtA)^{2}}{2\lambda dt}}| \phi,0\rangle.
\end{eqnarray}
\noindent We have written $dB'(t-ndt)=B'(t+dt-ndt)-B'(t-ndt)$. We have also written $P$ as the joint probability 
of $\{B'(t+dt),B'(t),..,B'(dt)\}$, instead of the joint probability of $\{dB'(t), dB'(t-dt), ...dB'(0)\}$, which we can do since the Jacobian 
determinant for the change of variables is 1. 

Now, we are interested in finding the probability  $P[B'(t+dt)]$, regardless of what Brownian path leads to the 
value of $B'(t+dt)$. To obtain  this, we integrate Eq.(\ref{18}) over all $B'$'s except $B'(t+dt)$.  These integrals are easily done,
since each  $B'(mdt)$ appears in just two (adjacent) gaussians in the product, and 
\begin{eqnarray}
&&\int_{-\infty}^{\infty}dB\frac{1}{\sqrt{2\pi c_{1} }}e^{-\frac{(B-a_{1})^{2}}{2c_{1}}}\frac{1}{\sqrt{2\pi c_{2} }}e^{-\frac{(B-a_{2})^{2}}{2c_{2}}}\nonumber\\
&&\qquad\qquad\qquad\qquad=\frac{1}{\sqrt{2\pi (c_{1}+c_{2} )}}e^{-\frac{(a_{1}-a_{2})^{2}}{2(c_{1}+c_{2})}}.\nonumber
\end{eqnarray}
\noindent The result for $P[B'(t+dt)]$ is

\begin{eqnarray}\label{19}
P&=&\frac{1}{\sqrt{2\pi\lambda (t+ dt)}}
\langle\phi,0|e^{-\frac{(B'(t+dt)-2\lambda (t+dt)A)^{2}}{2\lambda dt}}| \phi,0\rangle\nonumber\\
&=& \frac{1}{\sqrt{2\pi\lambda (t+ dt)}}e^{-\frac{(B'(t+dt))^{2}}{2\lambda dt}}\langle\phi,t+dt| \phi,t+dt\rangle.
\end{eqnarray}

\section{CSL}\label{secIII}

We have arrived at CSL's two equations in the form of the Schr\"odinger equation (\ref{13}) and the Probability Rule (\ref{19}). It is useful to incorporate the exponential factor in (\ref{19}) in the Schr\"odinger equation (and un-prime $B$, and replace $t+dt$ by $t$), so that its solution then becomes
\begin{equation}\label{20}
 |\phi,t\rangle=e^{-\frac{1}{4\lambda t}[B(t)-2\lambda t A]^{2}}|\phi,0\rangle, 
\end{equation}
\noindent leaving the Probability Rule in the simple form
\begin{equation}\label{21}
P[B(t)]dB(t)=\frac{dB(t)}{\sqrt{2\pi\lambda t}}\langle\phi,t| \phi,t\rangle.
\end{equation}

Eqs.(\ref{20}),(\ref{21}) define CSL. Everything that follows is based upon these two equations.

\subsection{Collapse mechanism}\label{secIIIA}
Let's see how they contrive to collapse the state vector. Suppose the initial state vector is 
\begin{equation}\label{22}
|\phi,0\rangle=\sum_{n} c_{n}|a_{n}\rangle.
\end{equation}
\noindent $|\phi,0\rangle$ is assumed normalized to 1, so $\sum_{n}| c_{n}|^{2}$=1.  Eqs.(\ref{20}), (\ref{21}) then become

\begin{equation}\label{23}
 |\phi,t\rangle=\sum_{n} c_{n}|a_{n}\rangle e^{-\frac{1}{4\lambda t}[B(t)-2\lambda ta_{n}]^{2}} 
\end{equation}
\noindent and 
\begin{equation}\label{24}
P[B(t)]dB(t)=\frac{dB(t)}{\sqrt{2\pi\lambda t}}\sum_{n} |c_{n}|^{2}e^{-\frac{1}{2\lambda t}[B(t)-2\lambda ta_{n}]^{2}}.
\end{equation}

Now, assume all the $a_{n}$ are unequal. Then,  Eq.(\ref{24}) describes a bunch of gaussians whose centers at $2\lambda ta_{n}$ drift further and further apart, while their widths  $\sqrt{\lambda t}$ spread much more slowly. As $t$ increases, the gaussians have less and less overlap.  

Then, for the set of $B(t)$'s which lie within the $m$th gaussian, the  state vector and probability are, to an excellent approximation (which becomes exact for $t\rightarrow\infty$),
\begin{equation}\label{25}
 |\phi,t\rangle\approx  c_{m}|a_{m}\rangle e^{-\frac{1}{4\lambda t}[B(t)-2\lambda ta_{m}]^{2}} 
\end{equation}
\noindent and 
\begin{equation}\label{26}
P[B(t)]dB(t)\approx\frac{dB(t)}{\sqrt{2\pi\lambda t}} |c_{m}|^{2}e^{-\frac{1}{2\lambda t}[B(t)-2\lambda ta_{m}]^{2}}.
\end{equation} 

For this set of $B(t)$'s, the integrated probability in (\ref{26}) is $ \approx |c_{m}|^{2}$, giving the Born Rule. 

 For any $B(t)$ in this set, the normalized state vector is 
\begin{equation}\label{27}
  |\psi,t\rangle\equiv \frac{|\phi,t\rangle}{\sqrt{\langle\phi,t|\phi,t\rangle}}\approx |a_{m}\rangle
  \end{equation}
\noindent giving the collapsed state vector .
\subsection{Refinements: Density Matrix}\label{secIIIB}

The density matrix, as usual, is constructed from the state vectors and their associated probabilities, here given by Eqs.(\ref{20}),(\ref{21}):
\begin{eqnarray}\label{28}
\rho(t)&\equiv&\int_{-\infty}^{\infty} P[B(t)]dB(t)\frac{|\phi,t\rangle\langle\phi,t|}{\langle\phi,t|\phi,t\rangle}\nonumber\\
&=&\int_{-\infty}^{\infty}\frac{dB(t)}{\sqrt{2\pi\lambda t}}\langle\phi,t| \phi,t\rangle
\frac{|\phi,t\rangle\langle\phi,t|}{\langle\phi,t| \phi,t\rangle}\nonumber\\
&&\negmedspace\negmedspace\negmedspace\negmedspace\negmedspace\negmedspace\negmedspace\negmedspace\negmedspace\negmedspace\negmedspace\negmedspace\negmedspace\negmedspace\negmedspace\negmedspace\negmedspace\negmedspace
=\int_{-\infty}^{\infty}\frac{dB(t)}{\sqrt{2\pi\lambda t}}  e^{-\frac{1}{4\lambda t}[B(t)-2\lambda t A]^{2}}|\phi,0\rangle\langle\phi,0| e^{-\frac{1}{4\lambda t}[B(t)-2\lambda t A]^{2}}\nonumber\\
&=&e^{-\frac{\lambda t}{2}[A_{L}-A_{R}]^{2}}|\phi,0\rangle\langle\phi,0|.
\end{eqnarray}
\noindent  In Eq.(\ref{28}), $A_{L},A_{R}$ mean that  these operators act on the left or the right respectively of the initial density matrix $|\phi,0\rangle\langle\phi,0|$.

With the initial density matrix (\ref{22}), the matrix elements of the density matrix at time $t$ is found from (\ref{28}) to be
\begin{equation}\label{29}
\langle a_{n}|\rho(t)|a_{m}\rangle=c_{n}c^{*}_{m}e^{-\frac{\lambda t}{2}[a_{n}-a_{m}]^{2}},
\end{equation}
\noindent showing how the off-diagonal elements decay while the diagonal elements remain constant, the collapse rate increasing as the eigenvalue differences increase. 

According to Eq.(\ref{28}), the differential equation satisfied by the density matrix is
\begin{equation}\label{30}
\frac{d\rho(t)}{dt}=-\frac{\lambda}{2}[A,[A,\rho(t)]],
\end{equation}
\noindent which is the simplest possible Lindblad equation.\cite{Lindblad}
\subsection{Refinements: Hamiltonian}\label{secIIIC}

To add the Hamiltonian to the state vector dynamics, consider the evolution over an infinitesimal time interval: it and the probability rule become
 \begin{subequations}
\begin{eqnarray}\label{31}
|\phi,t\rangle&=&e^{-iHdt-\frac{1}{4\lambda dt}[dB(t)-2\lambda dtA]^{2}}|\phi,t-dt\rangle\nonumber\\
&=&e^{-dt[iH+\frac{1}{4\lambda}[w(t)-2\lambda A]^{2}}|\phi,t-dt\rangle\label{31a}\\
P(w)dw&=& \frac{\langle \phi, t|\phi,t\rangle}{\langle \phi, t-dt|\phi,t-dt\rangle}\frac{dw(t)}{\sqrt{2\pi\lambda/dt}}\label{31b}
\end{eqnarray}
 \end{subequations}
 \noindent where $w(t)\equiv dB(t)/dt$ is called white noise.  

Over a finite time interval,   Eqs.(\ref{31a}),(\ref{31b}) imply

 \begin{subequations}
\begin{eqnarray}\label{32}
|\phi,t\rangle&=&{\cal T}e^{-\int_{0}^{t}dt'\{iH(t')+\frac{1}{4\lambda}[w(t')-2\lambda A]^{2}\}}|\phi,0\rangle\label{32a}\\
&&P(w)Dw= \langle \phi, t|\phi,t\rangle\prod_{t'=0}^{t-dt}\frac{dw(t')}{\sqrt{2\pi\lambda/dt}}\label{32b}
\end{eqnarray}
 \end{subequations}
\noindent where ${\cal T}$ is the time-ordering operator.  

To summarize, for each white noise function $w(t)$ there is a corresponding state vector $|\phi,t\rangle$ given by Eq.(\ref{32a}), one of which is supposed to be realized in nature with probability (\ref{32b}).  Generally, the hamiltonian evolution and the collapse-hamiltonian evolution compete against each other. This can give rise to effects which suggest experimental tests of the collapse theory vis-\`a-vis standard quantum theory/collapse postulate. 

`White noise' was named after the sound which has all frequencies in equal amounts,  in analogy to white light. It was named in a paper on the acoustics in airplanes,\cite{whitenoise} where the authors wrote:
 
 $\hbox{}$
 
 That white noise is annoying needs little argument. No one
has been found who really enjoys it.
 
  $\hbox{}$
  
 \noindent However, here it is enjoyed, in its role as the ``chooser" of the collapsed state.  
\subsection{Refinements: More Collapse-Generating Operators}\label{secIIID}
It is a straightforward generalization to describe collapse to a joint basis of operators $A^{\alpha}$ which commute, $[A^{\alpha},A^{\beta}]=0$. This requires one white noise function $w^{\alpha}$ for each $A^{\alpha}$.  The state vector evolution is
\begin{equation}\label{33}
|\phi,t\rangle={\cal T}e^{-\int_{0}^{t}dt'\{iH(t')+\frac{1}{4\lambda}\sum_{\alpha}[w^{\alpha}(t')-2\lambda A^{\alpha}]^{2}\}}|\phi,0\rangle
\end{equation}
\noindent and the corresponding  density matrix evolution is 
\begin{equation}\label{34}
\frac{d\rho(t)}{dt}=-i[H,\rho(t)]-\frac{\lambda}{2}\sum_{\alpha}[A^{\alpha},[A^{\alpha},\rho(t)]].
\end{equation}
The ensemble average of an operator ${\cal O}$ shall be denoted $\overline{\cal O}(t)\equiv \hbox {Tr}{\cal O}\rho(t)$, where $ \hbox {Tr}$ is the trace operation.  
Then, Eq.(\ref{34}) gives
\begin{equation}\label{35}
\frac{d\overline{\cal O}(t)}{dt}=-i\overline{[\cal{ O},H]}(t)-\frac{\lambda}{2}\sum_{\alpha}\overline{[A^{\alpha},[A^{\alpha},{\cal O}]]}(t).
\end{equation}
\section{Non-Relativistic CSL}\label{secIIIF}

Finally, here is the CSL proposal to describe the non-relativistic world.\cite{Pearle89, GPR}.   The index $\alpha$ in Eq.(\ref{33}) is changed to a continuum index ${\bf x}$, so 
the `chooser' $w({\bf x},t)$, rather than being a set of random functions, is a random field:
\begin{equation}\label{36}
|\phi,t\rangle={\cal T}e^{-i\int_{0}^{t}dt'H(t')-\frac{1}{4\lambda}\int_{0}^{t}dt'\int d{\bf x}'[w({\bf x}',t')-2\lambda A({\bf x}')]^{2}}|\phi,0\rangle.  
\end{equation}
\noindent The set of collapse-generating operators are mass-density operators,  `smeared' over a sphere of radius $a$:
\begin{equation}\label{37}
A({\bf x})\equiv\sum_{n}\frac{m_{n}}{M}\frac{1}{(\pi a^{2})^{3/4}}\int d{\bf z}e^{-\frac{1}{2a^{2}}[{\bf x}-{\bf z}]^{2}}\xi_{n}^{\dagger}({\bf z})\xi_{n}({\bf z}).
\end{equation}
\noindent Here, $\xi_{n}^{\dagger}({\bf z})$ is the creation operator for a particle of type $n$ at ${\bf z}$.  $m_{n}$ is the mass of this particle and $M$ is the mass of a nucleon (say, the neutron).  Thus, in 
ordinary matter, it is the nucleons  which are mostly responsible for collapse. Experimental results\cite{mass} have dictated that the effective collapse rate  in Eq.(\ref{36}) be mass-proportional, $\sim \lambda m_{n}$.

Assuming the theory is correct, the parameter values of $\lambda, a$ should be determined by experiment.\cite{Tumulka} Until then, we shall provisionally adopt the parameter values chosen by Ghirardi, Rimini and Weber\cite{GRW} in their instantaneous collapse theory, $\lambda\approx 
10^{-16}\hbox {sec}^{-1}, a\approx 10^{-5}\hbox {cm}$.  However,  it should be mentioned that Adler\cite{Adler} has given an argument for $\lambda$ 
to be as large as $\approx 10^{-11}\hbox {sec}^{-1}$.
 
The density matrix evolution equation (\ref{34}) becomes, using (\ref{37}),
\begin{eqnarray}\label{38}
\frac{d\rho(t)}{dt}&=&-i[H,\rho(t)]-\frac{\lambda}{2}\sum_{k,n}\frac{m_{k}m_{n}}{M^{2}}\cdot\nonumber\\
&&\negmedspace\negmedspace\negmedspace\negmedspace\negmedspace\negmedspace\negmedspace\negmedspace
\negmedspace\negmedspace\negmedspace\negmedspace\negmedspace\negmedspace\negmedspace\negmedspace
\frac{1}{(\pi a^{2})^{3/2}}\int d{\bf x}\int d{\bf z}\int d{\bf z}'e^{-\frac{1}{2a^{2}}[{\bf x}-{\bf z}]^{2}}e^{-\frac{1}{2a^{2}}[{\bf x}-{\bf z}']^{2}}\cdot\nonumber\\
&&[\xi_{k}^{\dagger}({\bf z})\xi_{k}({\bf z})[\xi_{n}^{\dagger}({\bf z}')\xi_{n}({\bf z}'),\rho(t)]]\nonumber\\
&=&-i[H,\rho(t)]-\frac{\lambda}{2}\sum_{k,n}\frac{m_{k}m_{n}}{M^{2}}\cdot\nonumber\\
&&\negmedspace\negmedspace\negmedspace\negmedspace\negmedspace\negmedspace\negmedspace\negmedspace
\negmedspace\negmedspace\negmedspace\negmedspace\negmedspace\negmedspace\negmedspace\negmedspace
\negmedspace\negmedspace\negmedspace\negmedspace\negmedspace\negmedspace\negmedspace\negmedspace
\int d{\bf z}\int d{\bf z}'e^{-\frac{1}{4a^{2}}[{\bf z}-{\bf z}']^{2}}[\xi_{k}^{\dagger}({\bf z})\xi_{k}({\bf z})[\xi_{n}^{\dagger}({\bf z}')\xi_{n}({\bf z}'),\rho(t)]].
\end{eqnarray}
\section{Free Small Clump}

In the rest of this paper, we shall illustrate CSL by discussing the force-free behavior of the center of mass (cm) of a small (dimensions$<a$) clump of ordinary matter.\footnote {The  internal excitation of the matter is not discussed here.  Collapse narrows wave packets,  resulting in atomic and nuclear `anomalous' excitation (i.e., collapse-generated, not predicted by standard quantum theory).\cite{mass}. Experimental limits on such excitation strongly suggests the effective mass-proportionality of the collapse rate, as we have mentioned.  Incidentally, it can be argued \cite{wenergy} that the increasing particle energy entails a concomitant  decrease in the $w$-field energy, so total energy is conserved.} 

For simplicity we shall neglect the electrons, and take there to be $N$ nucleons, regarded as a single type of particle, of mass $m=M$, which are very good approximations 
for our calculations.  Then, in the particle position basis, $|{\bf x}\rangle\equiv | {\bf x}_{1},....{\bf x}_{i},...{\bf x}_{N}\rangle$, using 
$\xi_{k}^{\dagger}({\bf z})\xi_{k}({\bf z})|{\bf x}\rangle=\sum_{i=1}^{N}\delta({\bf z}-{\bf x}_{i}) |{\bf x}\rangle$,            Eq.(\ref{38})
becomes
\begin{eqnarray}\label{39}
&&\frac{d\langle{\bf x}|\rho(t)|{\bf x}'\rangle}{dt}=-i\langle{\bf x}|[H,\rho(t)]|{\bf x}'\rangle-\langle{\bf x}|\rho(t)|{\bf x}'\rangle\cdot\nonumber\\
&&\negmedspace\negmedspace\negmedspace\negmedspace\negmedspace\negmedspace\negmedspace\negmedspace
\negmedspace\negmedspace\negmedspace\negmedspace\negmedspace\negmedspace\negmedspace\negmedspace
\frac{\lambda}{2}\sum_{i,j=1}^{N}\Big[e^{-\frac{1}{4a^{2}}[{\bf x}_{i}-{\bf x}_{j}]^{2}}+e^{-\frac{1}{4a^{2}}[{\bf x}'_{i}-{\bf x}'_{j}]^{2}}-2e^{-\frac{1}{4a^{2}}[{\bf x}_{i}-{\bf x}'_{j}]^{2}}\Big].
\end{eqnarray}

Define the center of mass coordinate ${\bf X}\equiv N^{-1}\sum_{i=1}^{N}{\bf x}_{i}$ and the relative coordinates 
${\bf y}_{i}\equiv{\bf x}_{i}-{\bf X}$.  Because it is a `small' clump, $\exp-[{\bf y}_{i}-{\bf y}_{j}]^{2}/4a^{2}\approx 1$ and  
$\exp-[{\bf X}+{\bf y}_{i}-{\bf X}'-{\bf y}'_{j}]^{2}/4a^{2}\approx \exp-[{\bf X}-{\bf X}']^{2}/4a^{2}$.  With the density matrix assumed to have the form of the direct product of cm and internal coordinate density matrices, we can take the trace over the internal coordinates in Eq.(\ref{39}) to obtain 
the equation for the evolution of $\langle{\bf X}|\rho(t)|{\bf X}'\rangle$.  

It is useful to express this equation in operator form,  writing the cm operator as $\hat {\bf X}$, and its conjugate cm momentum as $\hat {\bf P}$:
\begin{equation}\label{40}
\frac{d\rho(t)}{d t}=-i\Big[\frac{\hat {\bf P}^{2}}{2MN},\rho(t)\Big]
-\lambda N^{2}\Big[1-e^{-\frac{1}{4a^{2}}[\hat{\bf X}_{L}-\hat{\bf X}_{R}]^{2}}\Big]\rho(t).
\end{equation}

The associated state vector evolution equation (\ref{36}) is
\begin{equation}\label{41}
|\phi,t\rangle={\cal T}e^{-i\int_{0}^{t}dt'\frac{1}{2MN}\hat {\bf P}^{2}-
\frac{1}{4\lambda}\int_{0}^{t}dt'\int d{\bf x}'[w({\bf x}',t')-2\lambda NA({\bf x}')]^{2}}|\phi,0\rangle,  
\end{equation}
\noindent where
\begin{equation}\label{42}
A({\bf x}')\equiv\frac{1}{(\pi a^{2})^{3/4}}e^{-\frac{1}{2a^{2}}[{\bf x}'-\hat{\bf X}]^{2}}.
\end{equation}

To illustrate the use of the collapse part of Eq.(\ref{40}), 
consider the initial wave function
\begin{equation}\label{43}
|\phi, 0\rangle=\frac{1}{\sqrt{2}}\big[|L\rangle+|R\rangle\big], 
\end{equation}
\noindent where the states describe the clump to the left or right, with the two wave packet cm's separated by the distance D. Then (ignoring the kinetic energy),
\begin{equation}\label{44}
\frac{d\langle L|\rho(t)| R\rangle}{d t}=
-\lambda N^{2}\Big[1-e^{-\frac{D^{2}}{4a^{2}}}\Big]\langle L|\rho(t)| R\rangle.
\end{equation}
\noindent Thus, for $D>>a$, the collapse is described by exponential decay of the off-diagonal density matrix element with characteristic time 
$\lambda^{-1}=10^{16}$sec for a single nucleon.  For a $10^{-5}$cm cube of gold, where $N\approx 10^{8}$, the characteristic collapse time 
is $1/\lambda N^{2}=1$sec.  

	One might very well extend this theory to include massless particles by replacing mass-density of $A({\bf x})$ in Eq.(\ref{37}) by energy-density$/c^{2}$.  One might then regard it as holding in the co-moving frame of the universe\cite{co}, or as the limit of a relativistic CSL\cite{rel}.


\section{Collapse of a Packet}
We shall consider how a single wave packet undergoes collapse.
\subsection{Big Packet}
Consider a spread-out, real, positive, initial wave function such as
\begin{equation}\label{45}
\langle {\bf X}|\phi, 0\rangle=\frac{1}{(2\pi D^{2})^{3/4}}e^{-\frac{{\bf X}^{2}}{4D^{2}}},
\end{equation}
\noindent where $D>>a$.  We shall neglect the effect of the Hamiltonian.  We shall see that  the wave function collapses 
fairly rapidly to an approximately spherical wave function of size $a$, center location consistent with the Born Rule, and thereafter collapses 
more and more slowly to a smaller and smaller radius.  

First we calculate the ensemble average of the operator 
${\cal O}\equiv |{\bf X}\rangle\langle {\bf X}|$, so $\overline{{\cal O}}$ is the ensemble probability density at ${\bf X}$.  For any density matrix, it follows from the collapse part of Eq.(\ref{40}), using  Eq.(\ref{35}), that
\begin{equation}\label{46}
\frac{d\overline {|{\bf X}\rangle\langle {\bf X}|}}{d t}=
-\lambda N^{2}\langle {\bf X}|\Big[1-e^{-\frac{1}{4a^{2}}[\hat{\bf X}_{L}-\hat{\bf X}_{R}]^{2}}\Big]\rho(t)|{\bf X}\rangle =0.
\end{equation}
\noindent This, of course, doesn't say that collapse occurs but, if there is collapse, it says that the ensemble 
position probability distribution does not change from the initial distribution (Born Rule).

In order to see that there is indeed collapse, consider the ensemble average of the \textit{modular momentum}\cite{modmom} operator
${\cal O}\equiv \cos \hat{\bf  P}\cdot{\bf n}L$, where ${\bf n}$ is a unit vector pointing in some direction. This is 1/2 the sum of 
two operators, one which translates the wave function by distance $L$  in the ${\bf n}$ direction and the other in the $-{\bf n}$ direction.  Thus, its expectation value gives the overlap of the wave function with itself (all the wave functions are real and positive)  when translated.  For any density matrix, it follows from the collapse part of Eq.(\ref{40}) that
\begin{equation}\label{47}
\frac{d\overline { \cos \hat{\bf  P}\cdot{\bf n}L}}{d t}=
-\lambda N^{2}\Big[1-e^{-\frac{L^{2}}{4a^{2}}}\Big]\overline { \cos \hat{\bf  P}\cdot{\bf n}L} 
\end{equation}
\noindent Thus, for  $ L>> a$, the ensemble average of the overlap rate of the collapsing wave functions decreases as $\approx \lambda N^{2}$ but then it slows, e.g., for $L=a$, the collapse rate is $\approx .2\lambda N^{2}$. 
\subsection{ Small  Packet}\label{SP}
If the size of the wave function is less than $a$, one can utilize an approximate density matrix evolution equation obtained by expanding the exponential in Eq.(\ref{40}), retaining only the leading term:
\begin{equation}\label{48}
\frac{d\rho(t)}{d t}=-i\Big[\frac{\hat {\bf P}^{2}}{2MN},\rho(t)\Big]
-\frac{\lambda N^{2}}{4a^{2}}\sum_{1=1}^{3}[\hat{X}_{i},[\hat{X}_{i}, \rho(t)]]. 
\end{equation}
The state vector evolution which yields this density matrix evolution is\footnote{How can  Eq.(\ref{49}), where $w$ is just a function of $t$, arise from Eq.(\ref{40}), where $w$ is a field, depending upon ${\bf x}$ 
as well as $t$?  As far as I am aware, this has not been discussed before, so we treat it in Appendix \ref{B}.  More generally, it involves changing the collapse-generating operators $A^{\alpha}$ to a new, equivalent set, with concomitant change of white noise functions $w^{\alpha}(t)$ to a new, equivalent set.}  
\begin{eqnarray}\label{49}
|\phi,t\rangle&=&{\cal T}e^{-i\int_{0}^{t}dt'\frac{1}{2MN}\hat {\bf P}^{2}}\nonumber\\
&&\cdot e^{-\frac{1}{4\lambda}\int_{0}^{t}dt'\sum_{i=1}^{3}[w_{i}(t')-\sqrt{2}\lambda Na^{-1} \hat{X}_{i}]^{2}}|\phi,0\rangle.  
\end{eqnarray}

When the initial wave function is a gaussian, such as Eq.(\ref{45}) with $D<a$, since the Schr\"odinger equation is quadratic in ${\hat P}$ and ${\hat X}$, the solution is a gaussian. The exact solution to this problem can be found.\cite{RW}\cite{BassiP} 
We shall arrive at it here using the formalism we have presented. It suffices to solve the one-dimensional problem since, with initial wave function (\ref{45}), Eq.(\ref{49}) is the product of three terms, one for each dimension. 

We assume that the wave function at any time has the form
\begin{equation}\label{50}
\psi (X, t)=e^{-A(t)X^{2}+B(t)X+C(t)},
\end{equation}
\noindent and proceed to solve the Schr\"odinger equation which follows from the time derivative of (\ref{49}):
\begin{eqnarray}\label{51}
&&\frac{\partial}{\partial t}\psi (X, t)=\frac{i}{2m}\frac{\partial^{2}}{\partial X^{2}}\psi (X, t)\nonumber\\
&&\qquad-\Big[\frac{1}{4\lambda}w^{2}(t)-
\frac{\tilde\lambda}{\lambda}w(t)X+\frac{\tilde\lambda^{2}}{\lambda}X^{2}\Big]\psi (X, t),
\end{eqnarray}
  \noindent where $m\equiv NM$ and $\tilde\lambda\equiv\lambda N/\sqrt{2}a$. Inserting (\ref{50}) into (\ref{51}) we obtain 
 \begin{subequations}
\begin{eqnarray}\label{52}
\dot A&=&\frac{-2i}{m}A^{2}+\frac{\tilde\lambda^{2}}{\lambda},\label{52a}\\
\dot B&=&\frac{-2i}{m}AB+\frac{\tilde\lambda}{\lambda}w(t).\label{52b}
\end{eqnarray}
 \end{subequations}

Eq.(\ref{52a}) is a Ricatti equation, and can be solved by the ansatz $A=(m/2i)\dot F/F$.  It follows from (\ref{52a}) that $\ddot F=F(2i\tilde\lambda^{2}/m\lambda)$.
Thus, $F=\exp\pm\alpha(1+i)t$, where $\alpha\equiv\tilde\lambda/\sqrt{m\lambda }$ and
\begin{equation}\label{53}
A=\frac{m\alpha(1-i)}{2}\frac{e^{\alpha(1+i)t}-Ke^{-\alpha(1+i)t}}{e^{\alpha(1+i)t}+Ke^{-\alpha(1+i)t}}.
\end{equation}
\noindent where $K$ is a constant depending upon $D$.

We see that $A$, which characterizes the squared standard deviation of $X$, is the same for all $w(t)$. Thus, the wave function \textit{approaches an equilibrium size}, independently of its initial spread  $D$.  The equilibrium occurs because the Schr\'odinger evolution tends to spread the wave function while the collapse evolution tends to narrow it.  This takes place in characteristic time $\alpha^{-1}\approx 5\times 10^{4}/N$sec. The equilibrium spread in $X$ (its standard deviation) is $1/\sqrt{2(A+A^{*})}=1/\sqrt{2m\alpha}\approx 4/N^{1/2}$cm.

We shall henceforth assume either that the collapse process starts at negative times so that equilibrium is reached at time 0, or that $A$ initially has its equilibrium value.  Putting that value into (\ref{52b}) gives
\begin{equation}\label{54}
	\dot B=-\alpha(1+i)B+\frac{\tilde\lambda}{\lambda}w(t),  
\end{equation}
\noindent with solution 
\begin{equation}\label{55}
  B(t)=\frac{\tilde\lambda}{\lambda}\int_{0}^{t}dt'w(t')e^{-\alpha(1+i)(t-t')}.
\end{equation}
Knowing $A$ and $B$, the expectation values of position and squared position can be found from (\ref{50}):
\begin{subequations}
\begin{eqnarray}\label{56}
\langle X\rangle&\equiv&\frac{\langle \psi, t|X|\psi, t\rangle}{\langle \psi, t|\psi, t\rangle}=\frac{B+B^{*}}{2(A+A^{*})}\nonumber\\
&=&\frac{\alpha}{\tilde\lambda}\int_{0}^{t}dt'w(t')e^{-\alpha(t-t')}\cos\alpha(t-t'),\label{56a}\\
\langle X^{2}\rangle&\equiv&\frac{\langle \psi, t|X^{2}|\psi, t\rangle}{\langle \psi, t|\psi, t\rangle}=\langle X\rangle^{2}+\frac{1}{2(A+A^{*})}.
\label{56b}
\end{eqnarray}
\end{subequations}

To complete the solution, we need to find $C(t)$ but, since it is used to find the probability density $\langle \psi, t|\psi, t\rangle$, it is best that we calculate that directly from the Schr\"odinger equation:
\begin{eqnarray}\label{57}
\frac{d}{dt}\langle \psi, t|\psi, t\rangle&=&-\frac{w^{2}(t)}{2\lambda}\langle \psi, t|\psi, t\rangle+2\frac{\tilde\lambda}{\lambda}w(t)\langle \psi, t|X|\psi, t\rangle\nonumber\\
&&\qquad\qquad-2\frac{\tilde\lambda^{2}}{\lambda}\langle \psi, t|X^{2}|\psi, t\rangle\nonumber\\
&&=-\frac{1}{2\lambda}[w(t)-2\tilde\lambda\langle X\rangle]^{2}\langle \psi, t|\psi, t\rangle\nonumber\\
&&\qquad\qquad-\frac{\tilde\lambda^{2}}{\lambda}\frac{1}{A+A^{*}}\langle \psi, t|\psi, t\rangle.
\end{eqnarray}
\noindent Therefore, omitting the time-dependent factor arising from the last term of (\ref{57}) (which is absorbed in the normalization of the probability), and defining a new set of white noise functions 
\begin{eqnarray}\label{58}
v(t)&\equiv& w(t)-2\tilde\lambda\langle X\rangle\nonumber\\
&=&w(t)-2\alpha\int_{0}^{t}dt'w(t')e^{-\alpha(t-t')}\cos\alpha(t-t'),
\end{eqnarray}
\noindent the probability density is simply 
\begin{equation}\label{59}
\langle \psi, t|\psi, t\rangle=e^{-\frac{1}{2\lambda}\int_{0}^{t}dt'v^{2}(t')}.
\end{equation}
\noindent We note that $Dw=Dv$, since it follows from (\ref{58}) that the Jacobian of the transformation from $w$'s to $v$'s has 1's on the diagonal and 0's above the diagonal.  

In order to use (\ref{59}), it is necessary to obtain the inverse of the transformation (\ref{58}).  This can be done by taking the second 
derivative of  (\ref{58}), with the result
\begin{equation}\label{60}
\frac{d^{2}w(t)}{dt^{2}}=\frac{d^{2}v(t)}{dt^{2}}+2\alpha\frac{dv(t)}{dt}+2\alpha v(t).  
\end{equation}
\noindent Defining $v(t)$'s Brownian motion $\tilde B(t)$ by $v(t)=d\tilde B(t)/dt$. It then follows from (\ref{60}) that $w(t)$ can variously be written as
\begin{subequations}
\begin{eqnarray}\label{61}
w(t)&=&v(t)+2\alpha\int_{0}^{t}dt_{1}v(t_{1})\nonumber\\
&&\qquad\qquad+2\alpha^{2}\int_{0}^{t}dt_{1}\int_{0}^{t_{1}}dt_{2}v(t_{2}),\label{61a}\\
         &=&v(t)+2\alpha\int_{0}^{t}dt_{1}v(t_{1})[1+\alpha(t-t_{1})],\label{61b}\\
         &=&v(t)+2\alpha\tilde B(t)+2\alpha^{2}\int_{0}^{t}dt'\tilde B(t').
\end{eqnarray}
\end{subequations}

It then follows from the first equation in (\ref{58}) that $\langle X\rangle$ can be written as 
\begin{eqnarray}\label{62}
\langle X\rangle =\frac{1}{\sqrt{m\lambda}}\Big[\tilde B(t)+\alpha\int_{0}^{t}dt'\tilde B(t')\Big].   
\end{eqnarray}
\noindent One can then show, using (\ref{50}), (\ref{55}), (\ref{61})  and (\ref{62}), that  
\begin{eqnarray}\label{63}
\langle P\rangle =2iA\langle X\rangle-iB(t)=\frac{\tilde\lambda}{\lambda}\tilde B(t).
\end{eqnarray}

This problem is completely solved.  We see from Eqs.(\ref{62}),(\ref{63}) that, after the equilibrium packet size is achieved, the momentum expectation value undergoes Brownian motion and the position expectation value undergoes a motion that can be described as Brownian+.   

Any expectation value can be calculated, and any ensemble average expectation value can be calculated.  For example, although it can readily be found using the density matrix, the ensemble average of the squared position expectation value can be found from Eq.(\ref{56b}), using  $\langle X\rangle=(1/2\tilde\lambda)[w(t)-v(t)]$ (Eq.(\ref{58})), Eq.(\ref{61b}), and $\overline{v(t)v(t')}=\lambda\delta(t-t')$ (which follows from (\ref{59})): 
\begin{eqnarray}\label{64}
\overline{\langle X^{2}\rangle}&=&\frac{1}{2m\alpha}+\frac{1}{m\lambda}\overline{\Bigg[\int_{0}^{t}dt'v(t')[1+\alpha (t-t')]\Bigg]^{2}}\nonumber\\
&=&\frac{1}{2m\alpha}+\frac{1}{m}\int_{0}^{t}dt'[1+\alpha (t-t')]^{2}\nonumber\\
&=&\frac{1}{2m\alpha}+\frac{1}{m}[t+\alpha t^{2}+\alpha^{2}t^{3}/3].
\end{eqnarray}

 $\overline X^{2}\sim t$ behavior occurs for classical Brownian motion, modeled as a particle undergoing Newtonian dynamics with a random force and a viscous damping force.   In this case, the average Brownian `step' size is constant in time. 
 
$\overline X^{2}\sim t^{3}$ behavior occurs for classical Brownian motion when the viscous damping is removed.  This is essentially because the average Brownian `step' size
  increases with time. 
  
So, we have the picture of the final result of collapse, a wave packet of equilibrium size which undergoes classical random walk without viscous damping, with momentum generally increasing as it undergoes classical random walk. 
\section{Collapse of Interfering Packets}
It follows from the density matrix evolution Eq.(\ref{40}) that the interaction picture density matrix $\tilde\rho(t)\equiv U^{\dagger}(t) \rho(t)U(t)$ 
($U(t)\equiv\exp(- i\hat{\bf P}^{2}/2m)$) satisfies 
\begin{equation}\label{65}
\frac{d\tilde\rho(t)}{d t}=-\lambda N^{2}U^{\dagger}(t)\Big[1-e^{-\frac{1}{4a^{2}}[\hat{\bf X}_{L}-\hat{\bf X}_{R}]^{2}}\Big]U(t)\tilde\rho(t),
\end{equation}
\noindent with solution
\begin{equation}\label{66}
\tilde\rho(t)={\cal T}e^{-\lambda N^{2}\int_{0}^{t}dt'U^{\dagger}(t')\Big[1-e^{-\frac{1}{4a^{2}}[\hat{\bf X}_{L}-\hat{\bf X}_{R}]^{2}}\Big]U(t')}
\rho(0)
\end{equation}
\noindent or, going back to the density matrix $\rho(t)$, 
\begin{equation}\label{67}
\rho(t)={\cal T}e^{-\lambda N^{2}\int_{0}^{t}dt'U(t-t')\Big[1-e^{-\frac{1}{4a^{2}}[\hat{\bf X}_{L}-\hat{\bf X}_{R}]^{2}}\Big]U^{\dagger}(t-t')}
\rho_{0}(t).
\end{equation}
\noindent where $\rho_{0}(t)\equiv\exp (-iHt) \rho(0)\exp(iHt)$ is the density matrix without collapse. In the position representation, Eq.(\ref{67}) is
\begin{subequations}
\begin{eqnarray}\label{68}
\langle {\bf X}|\rho(t)|{\bf X}'\rangle&=&{\cal T}e^{-\lambda N^{2}\int_{0}^{t}dt'
\big[1-e^{-\frac{{\bf Z}^{2}(t-t')}{4a^{2}}}\big]}
\langle {\bf X}|\rho_{0}(t)|{\bf X}'\rangle,\nonumber\\\label{68a}\\
{\bf Z}(t-t')&\equiv&({\bf X}-\frac{t-t'}{mi}{\bf \nabla})-({\bf X}'+\frac{t-t'}{mi}{\bf \nabla}').\label{68b}
\end{eqnarray}
\end{subequations}

 We now note that, because $[X_{i}-X'_{i}, \nabla_{j}+\nabla'_{j}]=0$, it follows that $[{\bf Z}(t-t'),{\bf Z}(t-t'')]=0$, and so the time-ordering operation 
 ${\cal T}$ may be removed from Eq.(\ref{68a}). Also because this commutator vanishes, any product of powers of $Z$'s can be written in `normal-ordered form,' by which we mean that the $X$'s are to the left of the $\nabla$'s.  Denoting the normal ordered form by :  :, Eq.(\ref{68a}) becomes
 \begin{equation}\label{69}
\langle {\bf X}|\rho(t)|{\bf X}'\rangle=:e^{-\lambda N^{2}\int_{0}^{t}dt'
\big[1-e^{-\frac{{\bf Z}^{2}(t-t')}{4a^{2}}}\big]}:
\langle {\bf X}|\rho_{0}(t)|{\bf X}'\rangle
\end{equation}

We shall apply Eq.(\ref{69}) to the case where the uncollapsed density matrix $\rho_{0}(t)$ is constructed from a number of wave packets, 
\begin{equation}\label{70}
\langle {\bf X}|\rho_{0}(t)|{\bf X}'\rangle=\sum_{n,n'}c_{n}c_{n'}^{*}\phi_{n}({\bf X},t)\phi_{n'}^{*}({\bf X}',t).
\end{equation}
 The wave packets $\phi_{n}({\bf X},t)$ are to have well-defined momenta ${\bf k}_{n}({\bf X})$ at (almost) each point of the wave packet, which itself has dimensions large compared to the wavelength.  Thus, a wave packet could be a laboratory `plane wave,' a good approximation to an eigenstate of momentum ${\bf k}$.  It could be a cylindrical wave packet or a spherical wave packet  of momentum magnitude $k$ such as might be obtained by putting the `plane' wave packet through a slit or a circular hole. 
   
An important feature of such a packet $\phi_{j}({\bf X},t)$ is that 
\begin{equation}\label{71}
\langle{\bf X}|\hat{\bf P}|\phi_{n},t\rangle= \frac{1}{i}{\bf \nabla}\phi_{n}({\bf X},t)\approx {\bf k}_{n}({\bf X})\phi_{j}({\bf X},t)
\end{equation}
\noindent is a very good approximation.  Another important feature of such a packet is that (almost) each point in each wave packet can be considered as moving on a straight-line trajectory with constant velocity ${\bf k}_{n}({\bf X})/m$.  

Putting together Eqs.(\ref{68b}),(\ref{69}),(\ref{70}), we obtain for the ensemble's probability density at ${\bf X}$: 
\begin{eqnarray}\label{72}
&&\langle {\bf X}|\rho(t)|{\bf X}\rangle=\sum_{n,n'}c_{n}c_{n'}^{*}\phi_{n}({\bf X},t)\phi_{n'}^{*}({\bf X},t)\nonumber\\
&&\quad\cdot e^{-\lambda N^{2}\int_{0}^{t}dt'
\Big[1-e^{-\frac{1}{4a^{2}}\big[{\bf X}_{n}(t-t')-{\bf X}_{n'}(t-t')\big]^{2}}\Big]},
\end{eqnarray}
\noindent where
\begin{equation}\label{73} 
{\bf X}_{n}(t-t')\equiv{\bf X}-\frac{{\bf k}_{n}({\bf X})}{m}(t-t').   
\end{equation}
\noindent That is, consider a point on the  $n$th packet  which is located at ${\bf X}$ at time $t$.  Then, ${\bf X}_{n}(t-t')$ is the location that point 
had on the  $n$th packet at the earlier time $t'$. 

To summarize, we have seen in Eq.(\ref{40}) or (\ref{44}) that, when a clump is put into a superposition of two 
places with constant separation $D$, the two states play the gambler's ruin game, so that the off-diagonal elements 
of the density matrix decay at the rate $\lambda N^{2}[1-\exp-(D^{2}/4a^{2})]$.  Eq.(\ref{73}) says that, for a superposition 
of packets, the points on the packets, which end up at the same place ${\bf X}$ at time $t$, may be thought of as playing the 
gambler's ruin game with each other on the way to ${\bf X}$, with the above-mentioned distance-determining rate now varying with time, governing 
the collapse all along the way.  

Although it is not of concern here, we mention that, of course, the spatially separated points of a single packet, or of multiple packets, likewise mutually play the 
gambler's ruin game, and that description is obtained by considering the off-diagonal 
elements of the density matrix.  
\subsection{Mach-Zender Interference}

As is well known, the Mach-Zender interferometer has a rectangular shape, say, with half-silvered beam-splitters at the lower left and upper right corners, and fully-silvered mirrors at the other two corners.  An incoming wave packet splits into two equal packets at the first beam splitter. The packet going $\rightarrow$, $\uparrow$ has its sign reversed when it reflects at 90$^{o}$  from the front-surfaced mirror.  The packet going $\uparrow$, $\rightarrow$, $\uparrow$ gets no net sign change: one sign change at the first,  front-surfaced, beam splitter, one at the front-surfaced mirror, none at the second, back-surfaced, beam splitter. Thus, without collapse, there is no output in the $\uparrow$ direction.  

Although there are certainly velocity changes of the packets, they take place over a relatively brief time interval, so Eq.(\ref{72}) may be 
 applied seriatim.  Let $t$ be the time interval separating emergence from the two beam splitters.  At time $0$, the two packets start off with $c_{1}= c_{2}=1/\sqrt{2}$.  Thereafter, 
$|{\bf X}_{1}(t-t')-{\bf X}_{2}(t-t')|>>a$.  Moreover, the collapse rate is unaffected if a packet changes sign. Finally, at time $t$,   the second beam splitter has just made the amplitudes $c_{1}= c_{2}=1/2$ and, if packet 1 is 
$\phi_{1}({\bf X},t)$, then packet 2 is $-\phi_{1}({\bf X},t)$.  

Therefore, in Eq.(\ref{72}), since $\exp-[{\bf X}_{n} - {\bf X}_{n}]^{2}/4a^{2}=1$, $\exp-[{\bf X}_{1} - {\bf X}_{2}]^{2}/4a^{2}\approx 0$,    for a point ${\bf X}$ in the superposed wave packets, 
\begin{equation}\label{74}
\langle {\bf X}|\rho(t)|{\bf X}\rangle=\frac{1}{4}|\phi_{1}({\bf X},t)|^{2}2[1-e^{-\lambda N^{2}t}],
\end{equation}
\noindent and the probability that the clump emerges in the upward direction is
\begin{equation}\label{75}
P_{\uparrow}=\int d{\bf X} \langle {\bf X}|\rho(t)|{\bf X}\rangle=\frac{1}{2}[1-e^{-\lambda N^{2}t}].
\end{equation}
We see that, as time spent in the interferometer increases,  $P_{\uparrow}\rightarrow1/2$ since, asymptotically,  only one packet  survives to hit the second beam splitter, 
and that packet has equal likelihood of going $\uparrow$ or $\rightarrow$.  

\subsection{Two Slit Interference}

It should be clear that Eq.(\ref{72}) can be applied to any interference or diffraction problem.  Here we shall just consider the effect of collapse on the  
two-slit Fraunhofer interference pattern, 
with neglect of single-slit diffraction.  

Take the  two slits to be located at $x=\pm b$ and parallel to the $z$-axis. The pattern is observed at a point ${\bf X}$ on a screen located at  $y=L>>b$.  The vectors from 
the slits to the point on the screen are ${\bf r}_{1,2}\equiv  {\bf X}\pm{\bf i}b$.  Defining  $\theta$ as the angle ${\bf X}$ makes with the $y$-axis, and working only to first order in $\theta$, then ${\bf X}\approx L[{\bf i}\theta+{\bf j}]$, and $r_{1,2}\approx |{\bf X}|\pm b\theta$.  

The two cylindrical packets $\phi_{1,2}$ have wave number $k$, and are of equal amplitude $c_{1,2}=1/\sqrt{2}$ when they emanate from the two slits.   They reach the screen at point ${\bf X}$ at time $t$, where their amplitudes are $A\exp\pm ikb\theta$.  We are to consider the parts of the two  packets   traveling with speed $k/m$ on a straight line, each from its slit to the screen. Their separation  at any time $t'$ is $|{\bf X}_{1}(t-t')-{\bf X}_{2}(t-t')|=2b[1-(t'/t)]$.  Putting this into Eq.(\ref{72}) yields
\begin{eqnarray}\label{76}
&&\langle {\bf X}|\rho(t)|{\bf X}\rangle=\frac{A^{2}}{2}\Big[2+(e^{2ikb\theta}+e^{-2ikb\theta})\nonumber\\
		&&\qquad\qquad\qquad\qquad\cdot e^{-\lambda N^{2}[t-\int_{0}^{t}dt'e^{-\big(\frac{b}{a}\big)^{2}\big(1-\frac{t'}{t}\big)^{2}}]}\Big] \nonumber\\
		&=&A^{2}\Big[1+\cos(2kb\theta)
		 e^{-\lambda tN^{2}[1-\frac{a}{b}\int_{0}^{\frac{b}{a}}dve^{-v^{2}}]}\Big] \nonumber\\
&=& 2A^{2}\cos^{2}(kb\theta)  e^{-\lambda tN^{2}[1-\frac{\sqrt{\pi}a}{2b}\chi(b/a)]}\nonumber\\
&& \qquad\qquad\qquad+A^{2}\Big[1-e^{-\lambda tN^{2}[1-\frac{\sqrt{\pi}a}{2b}\chi(b/a)]}\Big]       
\end{eqnarray}
\noindent where $\chi(b/a)\equiv$ erf$(b/a)$.

Thus, we see that the two-slit two-packet interference pattern decays while the single packet non-interference pattern builds up as time increases. 
For $b>>a$,  the packet separation is $>>a$ for almost all the time and the collapse rate is $\lambda N^{2}$, as in the previous section.  For 
$b<<a$, the collapse rate is $\lambda N^{2}b^{2}/3a^{2}$. 

This concludes our discussion of free particle collapse dynamics. 

\appendix
\section{Proof That $R$ and $S$ must be diagonal}\label{A}
 We prove here that the real symmetric operators $R$ and $S$ in the Stratonovich Schr\"odinger equation for the un-normalized state vector, 
\begin{equation}\label{A1}
d|\phi,t\rangle=[RdB'+Sdt]|\phi,t\rangle
\end{equation}
\noindent must be diagonal in the $|a_{n}\rangle$ basis. This is in order that Eq.(\ref{A1}) give rise to the It\^o gambler's ruin condition Eq.(\ref{5}),
\begin{equation}\label{A2}
dx_{n}(t)=b_{n}({\bf x})dB(t).
\end{equation}

After putting Eq.(\ref{4}), $dB'=dB+fdt$, into Eq.(\ref{A1}), we convert that Stratonovich equation to an It{\^o}  equation, with the result
\begin{equation}\label{A3}
d|\phi,t\rangle=[RdB+Vdt]|\phi,t\rangle\hbox{ where }V\equiv S+Rf+\frac{\lambda}{2}R^{2}.
\end{equation}
\noindent We note that $V$ is also a real symmetric operator and, if we show $R$ and $V$ must be diagonal, then $S$ must also be diagonal.

Using the rules for manipulating It\^o equations, it is straightforward to find
\begin{subequations}
\begin{eqnarray}\label{A4}
&&d|\phi,t\rangle\langle \phi,t|=\{[RdB+Vdt],|\phi,t\rangle\langle \phi,t|\}\nonumber\\
&&\qquad\qquad\qquad\qquad +\lambda dtR|\phi,t\rangle\langle \phi,t|R, \\  
&&    d\langle \phi,t|\phi,t\rangle =2[ \langle \phi,t|R|\phi,t\rangle dB +  \langle \phi,t|V|\phi,t\rangle dt ]\nonumber\\
&&\qquad\qquad\qquad\qquad+\lambda dt\langle \phi,t|R^{2}|\phi,t\rangle\langle \phi,t|R,
\end{eqnarray}
\end{subequations}
\noindent where $\{M,N\}\equiv MN+NM$. Defining the density matrix $\rho(t)\equiv|\phi,t\rangle\langle \phi,t|/\langle \phi,t|\phi,t\rangle$ and $\overline{M}\equiv\hbox{Trace} M\rho$, 
we obtain from Eqs.(\ref{4}) 
and the  It\^o rules:
\begin{eqnarray}\label{A5}
d\rho&=&[\{R,\rho\}-2\rho\overline{R}]dB+dt[\{V,\rho\}-2\rho\overline{V}]+\nonumber\\
&&\lambda dt\Big[[R\rho R-\rho\overline{R^{2}}]-2\overline{R}[\{R,\rho\}-2\rho\overline{R}]\Big].
\end{eqnarray}

Now, $x_{n}(t)=\langle a_{n}|\rho(t)|a_{n}\rangle$. Thus, in order that the diagonal elements of 
 Eq.(\ref{A5}) agree with  Eq.(\ref{A2}), we see that the diagonal elements of Eq.(\ref{A5}) which do not 
	 multiply $dB$ must vanish for \textit{arbitrary} $\rho$:
 \begin{eqnarray}\label{A6}
&&0=[\{V,\rho\}_{nn}-2\rho_{nn}\overline{V}]+\lambda \Big[[(R\rho R)_{nn}-\rho_{nn}\overline{R^{2}}]\nonumber\\
&&\qquad\qquad\qquad\qquad-2\overline{R}[\{R,\rho\}_{nn}-2\rho_{nn}\overline{R}]\Big].
\end{eqnarray}
\noindent where $M_{nm}\equiv \langle a_{n}|M|a_{m}\rangle$

First, suppose that $\rho_{mm}=1$, where $m\neq n$, and all other matrix elements of $\rho$ vanish. It follows from Eq.(\ref{A6}) that 
\begin{equation}\label{A7}
0=(R\rho R)_{nn}=(R_{nm})^{2}.
\end{equation}
That is, all the off-diagonal elements of $R$ vanish, so $R$ is diagonal.

Second,  choose a density matrix for which $\rho_{nn}$, $\rho_{mm}=1-\rho_{nn}$, $\rho_{nm}$ do not vanish, but all other matrix elements of $\rho$ do vanish.  Then, using the diagonal nature of $R$, Eq.(\ref{A6}) may be written as
\begin{eqnarray}\label{A8}
&&0=2V_{nm}\rho_{nm}[1-2\rho_{nn}]\nonumber\\
&&+\rho_{nn}[1-\rho_{nn}]\Big[2(V_{nn}-V_{mm})\nonumber\\
&&+\lambda(R_{nn}-R_{mm})[(R_{nn}+R_{mm})\nonumber\\
&&-4(R_{nn}\rho_{nn}+R_{mm}(1-\rho_{nn}))\Big].
\end{eqnarray}
\noindent For fixed $\rho_{nn}$, a viable density matrix (non-negative eigenvalues which add up to 1) exists for $|\rho_{nm}|\leq\sqrt{\rho_{nn}(1-\rho_{nn})}$. But, as $\rho_{nm}$ 
is varied,  the first term in Eq.(\ref{A8}) varies while the rest of the terms remain fixed.  Thus, the first term must vanish, and this means that $V_{nm}=0$ for $n\neq m$, i.e., $V$ is diagonal as well as $R$.   
\section{Transformation of Operators and White Noise}\label{B}
Consider the general CSL form for the evolution of the state vector, Eq.(\ref{33}) 
\begin{equation}\label{B1}
|\phi,t\rangle={\cal T}e^{-\int_{0}^{t}dt'\{iH(t')+\frac{1}{4\lambda}\sum_{\alpha}[w^{\alpha}(t')-2\lambda A^{\alpha}]^{2}\}}|\phi,0\rangle
\end{equation}
\noindent We introduce a real orthonormal set of vectors $u_{\beta}^{\alpha}$, i.e., $\sum_{\alpha}u_{\beta}^{\alpha}u_{\beta'}^{\alpha}=\delta_{\beta\beta'}$,  $\sum_{\beta}u_{\beta}^{\alpha}u_{\beta}^{\alpha'}=\delta_{\alpha\alpha'}$.  Defining white noise functions $v^{\beta}(t)$ and complete commuting set of operators $Z^{\beta}$ by $w^{\alpha}(t)\equiv\sum_{\beta} u_{\beta}^{\alpha}v^{\beta}(t)$ and $A^{\alpha}(t)\equiv\sum_{\beta} u_{\beta}^{\alpha}Z^{\beta}(t)$, one readily sees that, in the exponent of  Eq.(\ref{B1}),
\begin{equation}\label{B2}
\sum_{\alpha}[w^{\alpha}(t)-2\lambda A^{\alpha}]^{2}=\sum_{\beta}[v^{\beta}(t)-2\lambda Z^{\beta}]^{2}.
\end{equation}
\noindent  The Jacobian of the transformation from $w$'s to $v$'s is 1 so, in using the Probability Rule (\ref{32b}), $Dw=Dv$.

We wish to apply such a transformation to Eqs.(\ref{41}), (\ref{42}) which, for simplicity, we limit to one-dimensional space:
\begin{subequations}
\begin{eqnarray}\label{B3}
&&|\phi,t\rangle={\cal T}e^{-i\int_{0}^{t}dt'\frac{1}{2MN}\hat {P}^{2}}\nonumber\\
&&\qquad\cdot e^{-\frac{1}{4\lambda}\int_{0}^{t}dt'\int dx'[w(x',t')-2\lambda NA(x')]^{2}}|\phi,0\rangle, \label{B3a}\\ 
&&A(x')\equiv\frac{1}{(\pi a^{2})^{1/4}}e^{-\frac{1}{2a^{2}}[x'-\hat{X}]^{2}}.\label{B3b}
\end{eqnarray}
\end{subequations}\label{B4}
\noindent We shall use as orthonormal functions the harmonic oscillator wave functions
\begin{equation}
u_{n}(x)\equiv C_{n}H_{n}(x/a)e^{-\frac{1}{2a^{2}}x^{2}}\medspace\hbox{where }\medspace C_{n}\equiv\frac{1}{\sqrt{\pi^{1/2}2^{n}n!a}}.
\end{equation}
\noindent With the definitions $v_{n}(t)\equiv\int dx w(x,t)u_{n}(x)$ and $\hat{Z}_{n}\equiv\int dx A(x)u_{n}(x)$, 
the exponent in Eq.(\ref{B3a}) may be written as
\begin{eqnarray}\label{B5}
&&-\frac{1}{4\lambda}\int_{0}^{t}dt'\int dx'[w(x',t')-2\lambda NA(x')]^{2}=\nonumber\\
&&\qquad\qquad-\frac{1}{4\lambda}\sum_{n=0}^{\infty}\int_{0}^{t}dt'[v_{n}(t')-2\lambda N\hat{Z}_{n}]^{2}.
\end{eqnarray}
\noindent Thus, we see how a white-noise field gets converted to an equivalent sum of white noise functions.

 Using the identity $\exp(-t^{2}+2tz)=\sum_{n=0}^{\infty}t^{n}H_{n}(z)/n!$, with $t\equiv\hat{X}/2a$, $z\equiv x'/a$, we find
\begin{eqnarray}\label{B6}
\hat{Z}_{n}&=&\int dx\frac{1}{(\pi a^{2})^{1/4}}e^{-\frac{1}{2a^{2}}[x'-\hat{X}]^{2}} u_{n}(x)\nonumber\\
&=&\frac{1}{C_{n}(\pi a^{2})^{1/4}}e^{-\frac{1}{4a^{2}}\hat{X}^{2}}\sum_{m=0}^{\infty}\frac{(\hat{X}/2a)^{m}}{m!}\int dx u_{m}(x)u_{n}(x)\nonumber\\
&=&e^{-\frac{1}{4a^{2}}\hat{X}^{2}}\frac{(\hat{X}/\sqrt{2}a)^{n}}{\sqrt{n!}}.
\end{eqnarray}
\noindent This leads to the density matrix evolution equation 
\begin{eqnarray}\label{B7}
&&\frac{d\rho(t)}{d t}=-i\Big[\frac{\hat {P}^{2}}{2MN},\rho(t)\Big]\nonumber\\
&&\negmedspace\negmedspace\negmedspace\negmedspace\negmedspace\negmedspace\negmedspace\negmedspace
-\frac{\lambda N^{2}}{2}\sum_{n=0}^{\infty}[e^{-\frac{1}{4a^{2}}\hat{X}^{2}}\frac{(\hat{X}/\sqrt{2}a)^{n}}{\sqrt{n!}},
[e^{-\frac{1}{4a^{2}}\hat{X}^{2}}\frac{(\hat{X}/\sqrt{2}a)^{n}}{\sqrt{n!}}, \rho(t)]]. \nonumber\\
\end{eqnarray}
\noindent If we expand $\exp-\hat{X}^{2}/4a^{2}$, we see that the $n=0$ term goes as $(\hat{X}/a)^{4}$  and the rest of the terms go as 
$(\hat{X}/a)^{n}$ to lowest order.  Therefore, the lowest order term comes from $n=1$. Upon neglect of the higher order terms, this  gives  
the density matrix evolution equation 
\begin{equation}\label{B8}
\frac{d\rho(t)}{d t}=-i\Big[\frac{\hat {P}^{2}}{2MN},\rho(t)\Big]
-\frac{\lambda N^{2}}{4a^{2}}[\hat{X},[\hat{X}, \rho(t)]]. 
\end{equation}
which is identical to the one-dimensional version of Eq.(\ref{48}).

\end{document}